\documentclass[%
 reprint,
 amsmath,amssymb,superscriptaddress,
 aps,
 prd
]{revtex4-2}

\usepackage{graphicx}
\usepackage{float}
\usepackage{physics}
\usepackage[dvipsnames]{xcolor}
\usepackage[hidelinks]{hyperref}
\usepackage[normalem]{ulem}
\usepackage{subcaption}
\usepackage{ragged2e}
\usepackage{dsfont}
\usepackage{ulem}

\usepackage{xcolor}

\begin{document}

\title{Is it worth the effort to find Lefschetz thimbles?\\ Integration contours with numerically optimal signal-to-noise \\\ ratios in simple fermionic toy models}

\author{Attila P\'asztor}
\affiliation{Institute  for Theoretical Physics, ELTE E\"otv\"os Lor\' and University, P\'azm\'any P\'eter s\'et\'any 1/A, H-1117 Budapest, Hungary}
\affiliation{MTA-ELTE Lendület ``Momentum'' Strongly Interacting Matter Research Group, Budapest, Hungary}
\author{D\'avid Peszny\'ak}
\email{Corresponding author: dpesznyak@student.elte.hu}
\affiliation{Institute  for Theoretical Physics, ELTE E\"otv\"os Lor\' and University, P\'azm\'any P\'eter s\'et\'any 1/A, H-1117 Budapest, Hungary}
\affiliation{MTA-ELTE Lendület ``Momentum'' Strongly Interacting Matter Research Group, Budapest, Hungary}
\affiliation{Department of Computational Sciences, HUN-REN Wigner Research Centre for Physics, Konkoly-Thege Miklós utca 29-33, H-1121 Budapest, Hungary}
\affiliation{Pennsylvania State University, Department of Physics, State College, Pennsylvania 16801, USA}

\date{\today}

\begin{abstract}
We perform a detailed analysis of the fermionic sign problem in a series of one-dimensional integrals, that are achieved as extreme (one-site) limits of genuine physics models. Altogether we studied a Hubbard-like, a Gross-Neveu-like, a Thirring-like and a Chern-Simons-like integral. 
We compare the Lefschetz-thimble structure for these integrals with contours obtained with the holomorphic flow equations at different flow times and with numerically optimized continuous integration contours, defined by a maximal value of the expectation values of the phases. With the holomorphic flow equation, we perform the large flow time limit, so that the average phase corresponds to its value on the thimbles. In all of these integrals we observe that the convergence to this value is not monotonic, meaning that there is an optimal flow time where the sign problem is weaker than it is on the thimbles. Furthermore, we find that for all of these toy models, numerical optimization can find continuous contours on which the sign problem is considerably weaker than it is both on the thimbles and at flowed integration contours at the optimal flow time.
\end{abstract}

\maketitle

\section{Introduction:} 
The numerical sign problem in fermionic path integrals represents one of the most significant challenges in computational physics. This is due to the fact that fermionic systems involve antisymmetric wave functions, and in the absence of extra symmetries (such as charge conjugation or particle-hole symmetry), Monte Carlo sampling often produces oscillating contributions with both positive and negative signs, leading to severe cancelations and an exponential degradation of the signal-to-noise ratio. This makes direct simulations of several physically interesting materials, such as high-temperature superconductors or the conjectured quark-gluon plasma cores of the heaviest neutron stars, practically impossible. Developing robust solutions or mitigations to the sign problem is therefore of central importance.

One class of strategies for addressing the sign problem is to deform the domain of integration into a complexified field space, where the oscillations of complex phases can be mitigated. For a recent review of these methods, see Ref.~\cite{Alexandru:2020wrj}. Among these methods, the Lefschetz thimble approach~\cite{Cristoforetti:2012su, Cristoforetti:2013wha, Fujii:2013sra, Alexandru:2015xva, Fujii:2015bua, DiRenzo:2015foa, DiRenzo:2017igr, Ulybyshev:2019fte, DiRenzo:2020cgp} has been particularly well studied: It decomposes the original real contour into a sum of steepest-descent manifolds (i.e., the thimbles) attached to critical points of the action. On each thimble, the imaginary part of the action is constant. While this approach is theoretically attractive, it does not completely solve the sign problem, as two sources of cancelations remain: (1) the contour deformation introduces a complex Jacobian, which itself produces fluctuating phases (a residual or local sign problem), and (2) in fermionic systems, the contributions of many thimbles must often be summed, leading to cancelations between them (a global sign problem). Moreover, identifying all relevant thimbles is a practically formidable task for all but the simplest fermionic theories~\cite{Kanazawa:2014qma, Tanizaki:2015rda}.

A different (but related) complexification approach is the use of the complex Langevin equation~\cite{Parisi:1983mgm, Damgaard:1987rr, Aarts:2008rr}. Here field space is not deformed, but rather enlarged into the full complexified field space. However, there are some serious issues with the applicability of the complex Langevin approach to fermionic sign problems~\cite{Mollgaard:2013qra, Aarts:2017vrv, Seiler:2023kes}. This is not surprising, as the formal justification for the correctness of the approach breaks down near zeros of the fermionic determinant~\cite{Aarts:2009uq, Aarts:2011ax}. It also turns out that there is a strong overlap between the regions of complexified field space explored in the complex Langevin and thimble approaches but with some important differences, discussed in Refs. ~\cite{Aarts:2014nxa,Boguslavski:2024zqf}. We are not going to discuss this connection further in this manuscript, and focus only on contour deformation approaches instead.

A lot of research has been focused on the practical issue of finding more efficient ways to identify the thimbles or at least to approximate them. In Ref.~\cite{Alexandru:2015sua}, the holomorphic flow equation was proposed. In this approach, a fictitious flow time $t$ is introduced, and the integration manifold is continuously deformed as $t$ increases. In the large-$t$ limit, the deformed contour asymptotically approaches the union of the relevant thimbles. Which thimbles are relevant and also the direction of integrating over them depend on the initial contour. While this method provides a systematic path toward the thimbles, in practice large flow times often lead to a reemergence of the original thimble approach: the dominant contributions of the different thimbles emerge by flowing small disjoint parts of the original integration contour, leading to ergodicity or overlap problems. Later, Ref.~\cite{Tanizaki:2017yow} proposed a different flow equation to generate integration manifolds. These flow equations avoid blow up at finite flow time, but as they also tend to the thimbles, the ergodicity problem remains.
To mitigate these problems, several approaches have been proposed~\cite{Fukuma:2017fjq, Fukuma:2020fez}. The common feature of these approaches is to use information from different flow times (e.g. by parallel tempering in the flow time parameter). Since smaller flow times do not yet have the mentioned ergodicity problems, they can allow for transitions between the regions around the different Lefschetz thimbles.

Another practical problem that has received considerable attention is the costly computation of the complex Jacobian associated with thimble-based approaches~\cite{Alexandru:2016lsn}. One promising strategy that has emerged is the use of machine learning techniques: a model can be trained to approximate the solution of the holomorphic flow equations at a fixed flow time~\cite{Alexandru:2017czx, Wynen:2020uzx}. If the machine learning model has a tractable Jacobian, this can dramatically reduce the computational cost of holomorphic flow-based calculations.

While these practical questions are important, there is a more fundamental issue that has received almost no attention in the literature: How close are the Lefschetz thimbles to the numerically optimal contours? In other words, are the Lefschetz thimbles truly the contours that maximize the signal-to-noise ratio? In this work, we set out to investigate this issue, using some simple fermionic toy models. There is only one paper in the literature we are aware of that looked into this issue, for one specific toy model: the heavy-dense limit of the Thirring model, which reduces to a one-dimensional integral. In this case, a simple two-parameter ansatz was shown to outperform the Lefschetz thimble approach~\cite{Lawrence:2018mve}. The notion of Lefschetz thimbles not being optimal is also implicitly implied in Refs. \cite{Lawrence:2023sfc, Lawrence:2021izu}. We are going to study this issue in several different models, to see if the nonoptimality of Lefschetz thimbles is a generic feature, and not just a peculiarity of that one integral in question.

The models we are going to study are all one-dimensional integrals derived from extreme limits of the Euclidean path integrals of some real physical systems. We consider integrals both with noncompact and compact integration domains. We also consider several physical sources for the sign problem: a chemical potential, a complex coupling and a $\theta$ term. We also consider both cases where the imaginary part of the Lefschetz thimbles remains finite, as well as a case where it diverges. By looking at models with several different features, we hope to arrive at a more generic picture. Altogether we study four models:
a Hubbard-like, a Gross-Neveu-like, a Thirring-like, and a Chern-Simons-like integral. For all of these models, finding the critical points of the action is straightforward, and the differential equations defining the thimbles can be solved numerically. Moreover, the holomorphic flow equations can also be integrated numerically to very high flow times, effectively reaching the large-$t$ limit. Finally, a simple optimization procedure can be applied to all of these integrals to identify the numerically optimal continuous contours, where the imaginary part of a contour element is a single-valued function of its real part. 

A key ingredient of this discussion is the definition of what we mean by an optimal contour. Through this work, what we mean by this is the maximum of the ratio of full-to-phase-quenched partition functions, which is equivalent to the expectation value of the sign in the phase-quenched theory,
\begin{align}
\label{eq:average_phase}
\frac{\mathcal{Z}}{\mathcal{Z}_{\mathrm{pq}}} = \left\langle e^{i \theta}\right\rangle_{\mathrm{pq}}\equiv\sigma\;.
\end{align}
This ratio is often used in the literature to measure the severity of the sign problem. While it is not mathematically guaranteed that optimizing this particular ratio will lead to the best signal-to-noise ratio for every observable, it is a reasonable choice, as it measures the severity of phase fluctuations, which is the main reason for the large noise in the first place. For signal-to-noise studies focusing on specific observables see Refs. \cite{Detmold:2020ncp, Detmold:2021ulb, Lin:2023svo}. We find that, in all cases, the numerically optimal contour exhibits a sign problem significantly weaker than that of the Lefschetz thimbles. Furthermore, we find that the convergence of the severity of the sign problem using holomorphic flow equations is not monotonic. There is a finite flow time at which the sign problem is considerably weaker than it is on the Lefschetz thimbles. Encouragingly, the numerically optimized contours do not tend to show the features of Lefschetz thimbles that make their numerical implementation unstable: unlike the Lefschetz thimbles, the numerically optimal contours show no cusps and no divergences to infinity. Thus, a strong focus on the Lefschetz thimble structure of different theories is not necessarily the best numerical approach to mitigating the sign problem.

\section{Methods}

\subsection{Sign problem}
In the Euclidean path integral representation the partition function is
\begin{align}
\label{eq:partition_fucntion}
    \mathcal{Z}=\int\mathcal{D}\phi\;e^{-S[\phi]}=\int\mathcal{D}\phi\;w[\phi]\;,
\end{align}
with $\phi$ denoting the spacetime dependent fields in the given theory, and $S$ is the Euclidean action functional. The weight $w[\phi]$ stands for the Boltzmann weight $e^{-S[\phi]}$, governing the probabilities of field configurations. The integral in Eq. \eqref{eq:partition_fucntion} is formally infinite dimensional, however, it can be made finite dimensional and well defined by discretizing spacetime onto a finite lattice. The expectation value of some observable $O$ with respect to the weight $w$ is defined as
\begin{align}
    \langle O\rangle=\frac{1}{\mathcal{Z}}\int\mathcal{D}\phi\;O[\phi]w[\phi]\;.
\end{align}
As long as $w[\phi]\in\mathbb{R}^+$, i.e., the action is real, well-established Markov chain Monte Carlo methods based on importance sampling are applicable to evaluate expectation values of different observables. Once the action develops a nonzero imaginary part, the weight $w[\phi]$ cannot be interpreted as a joint probability density of field configurations anymore, and importance sampling breaks down; this is called the complex action problem. In principle, the complex action problem can be bypassed via reweighting from a simulable theory, with weights $r[\phi]\in\mathbb{R}^+$. Our choice for these new weights is going to be $r=|w|$, i.e., the phase-quenched theory. Via reweighting, the expectation value of $O$ in the original theory can be expressed as a ratio of expectation values in the phase-quenched theory:
\begin{align}
\label{eq:reweighting}
    \langle O\rangle=\frac{\langle Oe^{i\theta}\rangle_{\mathrm{pq}}}{\langle e^{i\theta}\rangle_\mathrm{pq}}\;,
\end{align}
with $w/|w|=e^{i\theta}$, a pure phase, and $\langle\cdot\rangle_\mathrm{pq}$ notes taking expectation values with the weights $|w|$. This way the complex action problem is reduced to the sign problem, practically manifesting in large phase fluctuations, leading to nontrivial cancelations and exponentially large uncertainties in the number of degrees of freedom of the system considered. Note that the denominator in Eq. \eqref{eq:reweighting} is the average sign $\sigma$, defined in Eq. \eqref{eq:average_phase}. For $\sigma\approx1$ the sign problem is mild, while for $\sigma$ close to zero the sign problem is severe.

\subsection{Contour deformations}

    One can search for phase-quenched theories closer to the original theory by complexifying and deforming the original integration manifold $\mathcal{M}$. As long as the path integral weight $w[\phi]$ is a holomorphic function of the field variables, and the deformed manifold ${\mathcal{M}}'$ is in the same homology class as the original one, the multivariate generalization of Cauchy's theorem for complex integrals ensures that the partition function $\mathcal{Z}$ remains invariant, i.e, physics stays the same. However, the phase-quenched weight $|w|$ is not a holomorphic function, therefore the phase-quenched partition function $\mathcal{Z}_\mathrm{pq}$ does change upon deforming the integration manifold. Taking advantage of this, one might look for deformations that decrease $\mathcal{Z}_\mathrm{pq}$, consequently making $\sigma$ larger and the sign problem milder \cite{Cristoforetti:2012su, Alexandru:2020wrj}. It is convenient to parametrize the deformed manifold with the original one as
    \begin{align}
    \label{eq:contour_deformation}
    \begin{split}
        \mathcal{Z}&=\int_\mathcal{M}\mathcal{D}\phi\;w[\phi]=\int_{{\mathcal{M}}'}\mathcal{D}{\phi}'\;w[{\phi}']\\
        &=\int_{\mathcal{M}}\mathcal{D}\phi\;\det\mathcal{J}(\phi)w[{\phi}'(\phi)]=\int_\mathcal{M}\mathcal{D}\phi\;e^{-S_\mathrm{eff}[\phi]}\;,
    \end{split}
    \end{align}
     introducing the Jacobian of the contour deformation, $\mathcal{J}=\partial{\phi}'/\partial \phi$, and the shorthand notation $S_\mathrm{eff}$. When calculating the average phase $\sigma$ the effect of the Jacobian should also be taken into account as
     \begin{align}
         \sigma=\bigg\langle\frac{\det\mathcal{J}}{|\det\mathcal{J}|}e^{-i\mathrm{Im}S}\bigg\rangle_\mathrm{pq}^\mathrm{def}\;,
     \end{align}
     where $\langle\cdot\rangle_\mathrm{pq}^\mathrm{def}$ means that the expectation value is taken with respect to weights $|e^{-S_\mathrm{eff}}|$ with $S_\mathrm{eff}$ defined in Eq. \eqref{eq:contour_deformation}.
     There are numerous ways to define and generate complex contour deformations $\mathcal{M}\to{\mathcal{M}}'$ that ameliorate the sign problem. We briefly introduce the three methods considered by us in the following three subsections.
     
\subsection{Lefschetz thimbles}
\label{sec:Lefschetz}

    Lefschetz thimbles \cite{Cristoforetti:2012su} can be regarded as generalized steepest-descent contours, characterized by the property that the imaginary part of the action $S$ remains constant along them, hence phase fluctuations are absent when looking at individual thimbles.
    Thimbles are attached to critical points of the action $\phi_\mathrm{c}$, i.e., 
    \begin{equation}
    \partial S/\partial\phi|_{\phi_\mathrm{c}}=0\rm,
    \end{equation}
    and are defined as a set of initial conditions of the downward flow equation
    \begin{align}
    \label{eq:flow_equation}
        \frac{\mathrm{d}\phi'}{\mathrm{d}t}=-\bigg(\frac{\partial S}{\partial\phi'}\bigg)^*\;,
    \end{align} such that $\phi'(t\to\infty)=\phi_\mathrm{c}$, 
    where $(\cdot)^*$ denotes complex conjugation. This means that flows emanating from thimbles approach the critical point as the fictitious flow time $t$ is sent to infinity. Analogously, one can define dual thimbles as set of initial conditions for which the upward flow [same as Eq. \eqref{eq:flow_equation} but with the opposite sign] asymptotically goes to $\phi_\mathrm{c}$. Usually, there is more than one critical point, and a disjoint thimble $\mathcal{T}_a$ (and dual thimble $\mathcal{K}_a$) can be attached to each critical point $\phi_\mathrm{c}^a$ of the action, with $a$ indexing the critical points. Thus, the path integral can be reformulated as a linear combination of integrals evaluated on the different thimbles,
    \begin{align}
    \label{eq:thimbles}
        \mathcal{Z}=\int_\mathcal{M}\mathcal{D}\phi\;e^{-S[\phi]}=\sum_an_ae^{-i\mathrm{Im}S[\phi_\mathrm{c}^a]}\int_{\mathcal{T}_a}\mathcal{D}\phi'\;e^{-\mathrm{Re}S[\phi']}\;,
    \end{align}
    with the coefficients $n_a\in\mathbb{Z}$ determined by the number of intersections of $\mathcal{M}$ and $\mathcal{K}_a$, and their relative orientation. This way the constant phases can be pulled out of the integrals for each contributing thimble in Eq. \eqref{eq:thimbles}. However, the sign problem is not generally eliminated, only made less severe. A global sign problem remains if the constant phases $e^{-i\mathrm{Im}S[\phi^a_\mathrm{c}]}$, associated to the thimbles, produce cancelations between the thimbles. Furthermore, a complex Jacobian $\mathcal{J}=\partial \phi'/\partial\phi$, possibly different for each thimble, brings about a local sign problem as well (often called the residual sign problem). Hence the thimble decomposition could really solve the sign problem only if a single thimble contributes (or all contributing thimbles have the same phase), and the corresponding Jacobian can be ignored.

\subsection{Holomorphic flow equations}
\label{sec:holomorphic}

    Another method for constructing a complex deformation of the integration manifold that mitigates the sign problem is through the holomorphic flow equations \cite{Alexandru:2020wrj}, defined by
    \begin{align}
    \label{eq:holomorphic_flow_equation}
        \frac{\mathrm{d}\phi'}{\mathrm{d}t}=\bigg(\frac{\partial S}{\partial\phi'}\bigg)^*\;,
    \end{align}
    i.e., the upward flow version of Eq. \eqref{eq:flow_equation}, but with initial conditions $\phi'(t=0)=\phi$. The flowed manifold $\mathcal{M}_T$ at flow time $t=T$ is obtained by integrating Eq. \eqref{eq:holomorphic_flow_equation} from $t=0$ up to $T$, during which the field variables are continuously deformed from $\phi$ to $\phi'(T)$. In the limit $T\to\infty$, the flowed manifold tends toward the corresponding ensemble of Lefschetz thimbles. It can be readily shown that along the flow
    \begin{align}
        \frac{\mathrm{d}}{\mathrm{d}t}\mathrm{Re}S\geq0\qquad\text{and}\qquad\frac{\mathrm{d}}{\mathrm{d}t}\mathrm{Im}S=0\;,
    \end{align}
    resulting in that the phase oscillations originating from $e^{-S}$ get more and more suppressed as the holomorphic flow equations are evolved. Also note that the phase-quenched partition function also decreases as $\mathrm{Re}S$ increases along the flow, thus the average phase is expected to increase as well. However, this remains true only if the Jacobian of the deformation can be ignored. 
    
    The Jacobian of the deformation is calculated via another flow equation as
    \begin{align}
    \label{eq:Jacobian_flow}
        \frac{\mathrm{d}\mathcal{J}}{\mathrm{d}t}=(\mathcal{HJ})^*\;,
    \end{align}
    with $\mathcal{H}$ being the Hessian of the action with respect to the field variables. The initial condition for the Jacobian flow equation is $\mathcal{J}(t=0)=\mathds{1}$. The reliable calculation of the Jacobian is generally the most computationally demanding part of this method \cite{Alexandru:2016lsn}. 
    
    It is again essential to point out that for a complex Jacobian the monotonically increasing behavior of $\mathrm{Re}S$ does not guarantee that the sign problem gets weaker along the flow. Based on how $\mathcal{J}$ evolves with increasing flow time $t$, nontrivial cancelations can come into play again due to the local sign problem. The interplay along the flow of the two sign problems originating from $e^{-S}$ and $\mathcal{J}$ is crucial, and, similarly to the Lefschetz thimble scenario, only the former can be mitigated via the holomorphic flow method.

    \begin{figure*}[t]
        \centering
        \includegraphics[height=0.33\linewidth]{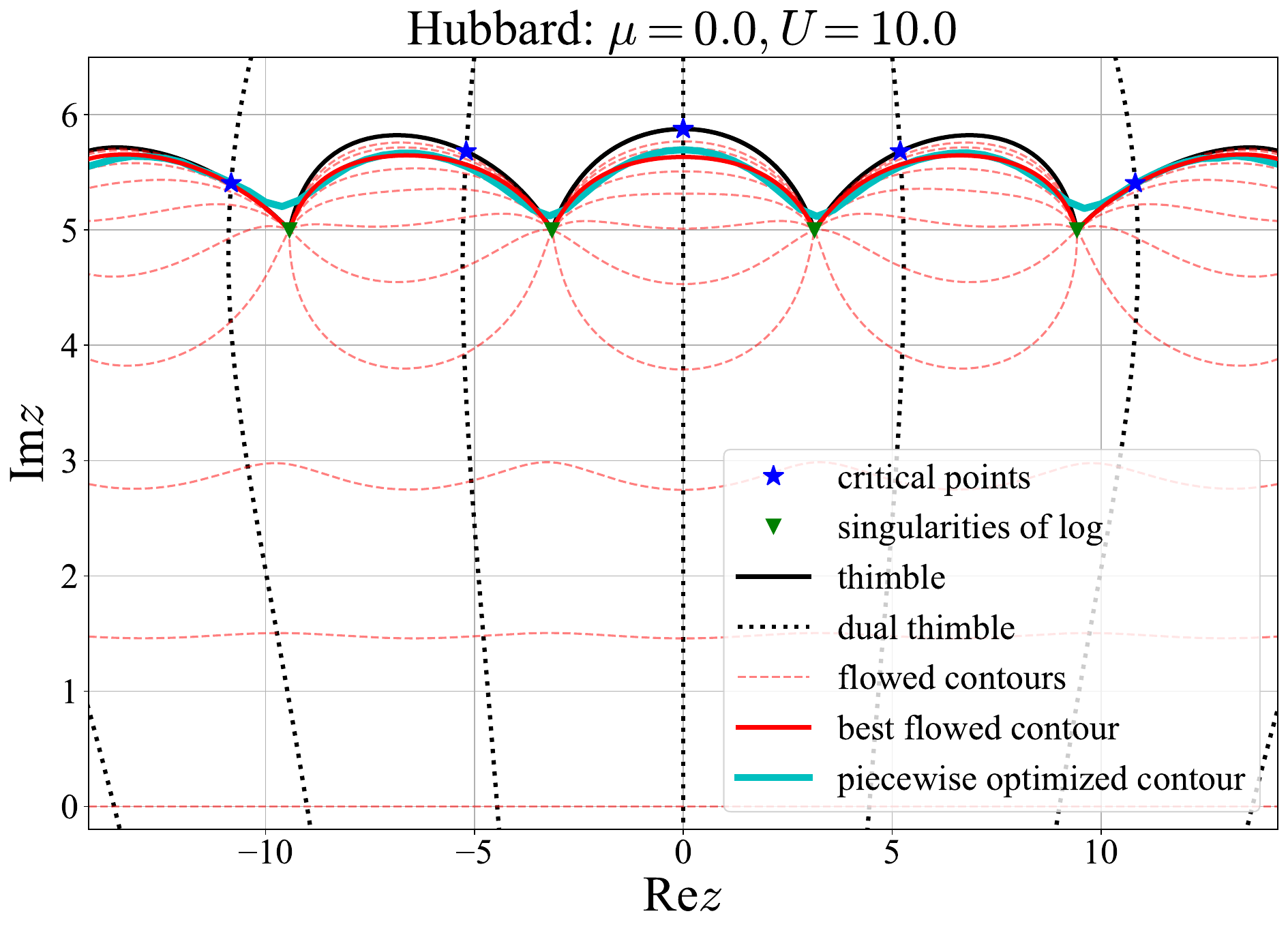}
        \includegraphics[height=0.33\linewidth]{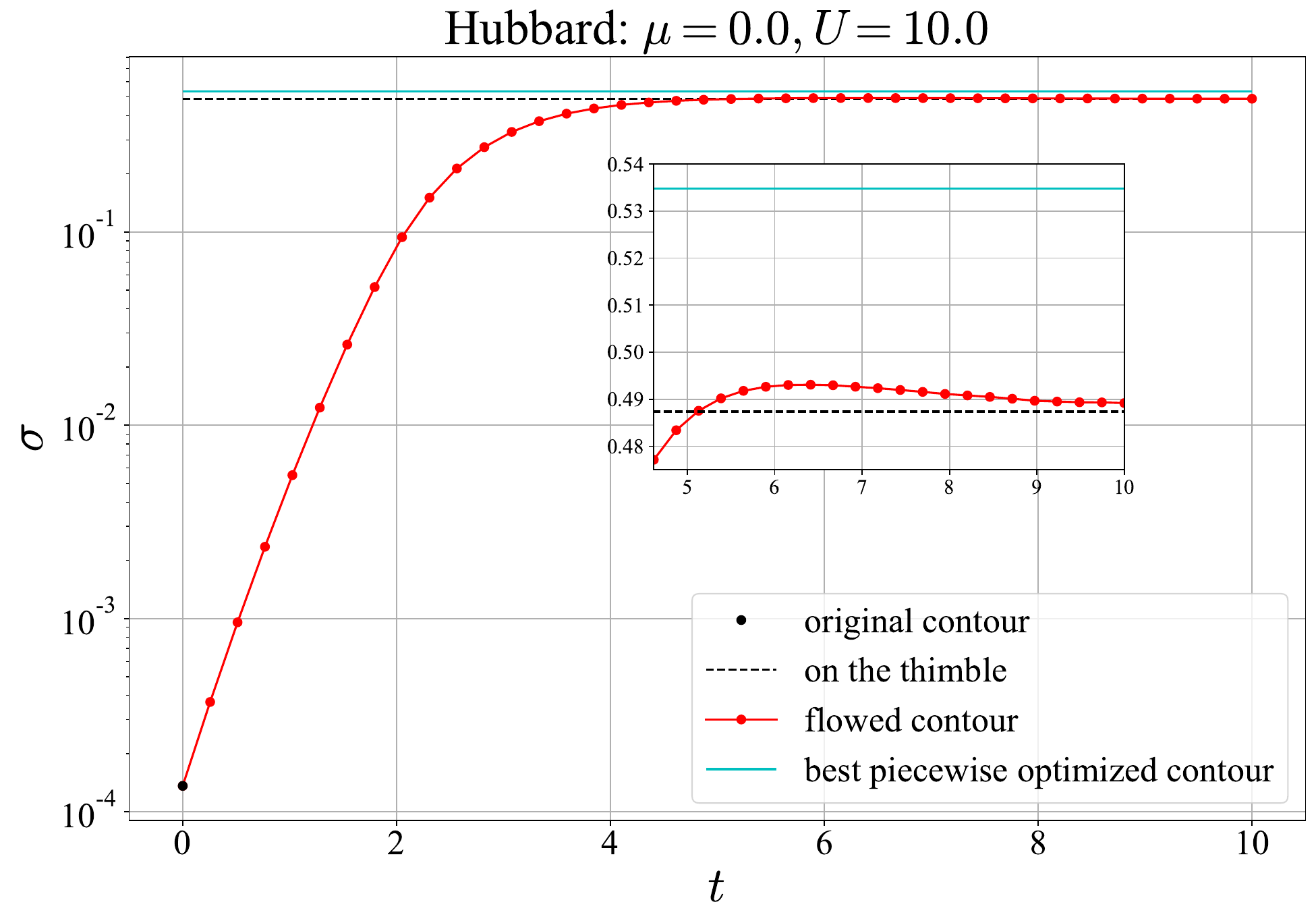}
        \caption{Left panel: Critical (blue stars) and singular (green wedges) points of the action of the one-site Hubbard model. The solid black lines are the thimbles, while the dotted black lines are the dual thimbles. The dashed red lines correspond to fixed flow time contours, while the solid red line is the one flowed contour that produces the maximal average sign. The solid cyan line is the optimum contour based on the sign optimization approach. Right panel: The evolution of the average sign $\sigma$ along the holomorphic flow (red dots), and its value on the thimble (black dashed line) and on the optimized contour (solid cyan line). Note that the vertical axis is in log-scale, with linear scale in the inset.}
        \label{fig:Hubbard}
    \end{figure*}

\subsection{Numerically optimal single-valued continuous contours}
\label{sec:optimization}

    In this paper we mainly focus on extreme limits of physics-motivated theories resulting in one-dimensional path integrals. To this end, we will continue with formulas restricted to such cases, and address multidimensional scenarios briefly later on. All arguments and equations so far can be trivially reduced to their one-dimensional limits.

    The partition function as a one-dimensional path integral reads
    \begin{align}
    \label{eq:1D}
    \begin{split}
        \mathcal{Z}&=\int_\mathcal{M}\mathrm{d}\varphi\;e^{-S(\varphi)}=\int_\mathcal{C}\mathrm{d}z\;e^{-S(z)}\\
        &=\int_\mathcal{M}\mathrm{d}\varphi\;\mathcal{J}(\varphi)e^{-S(z(\varphi))}=\int_\mathcal{M}\mathrm{d}\varphi\;e^{-S_\mathrm{eff}(\varphi)}\;,
    \end{split}
    \end{align}
    where the integration manifold $\mathcal{M}$ is simply $\mathbb{R}$, or some compact subset of it. The path integral weight $e^{-S(\varphi)}$ is a holomorphic function, and $\mathcal{C}$ denotes a curve that was continuously deformed into the complex plane starting from the initial manifold $\mathcal{M}$, without crossing any singularities (including singularities at infinity). Note that now the Jacobian is $\mathcal{J}=\partial z/\partial\varphi\in\mathbb{C}$. One can look for a contour deformation $\mathcal{M}\to\mathcal{C}$ that brings forth a milder sign problem by parametrizing the integration manifold with a finite set of parameters $\{p\}$, and then tune those parameters to produce a curve $\mathcal{C}(\{p\})$ on which the average phase $\sigma=\sigma(\{p\})$ is maximal. This method is generally referred to as sign (or path) optimization in the literature \cite{Mori:2017pne, Bursa:2018ykf,Bursa:2021org,Kashiwa:2018vxr,Mori:2019tux,Alexandru:2018fqp,Alexandru:2018ddf,Kashiwa:2020brj,Giordano:2023ppk,Tulipant:2022vtk,Lawrence:2025rnk}, and it is easily generalized to higher-dimensional integrals. The parameter tuning is performed using well-established machine learning techniques by minimizing a cost function. The cost function we used in this study is
    \begin{align}
    \label{eq:cost function}
        \mathcal{F}(\{p\})=-\log\sigma(\{p\})=-\log\mathcal{Z}+\log\mathcal{Z}_\mathrm{pq}(\{p\})\;.  
    \end{align}
    Note that the phase-quenched partition function depends on the deformation parameters $\{p\}$, while the original one does not (and must not). It is practical to set up the deformation such that it reproduces the original integration contour when all deformation parameters are set to zero. One needs the calculate gradients of the cost function with respect to the deformation parameters in order to use an optimizer  and the form in Eq. \eqref{eq:cost function} is beneficial in that regard because the gradients can be expressed as expectation values in the deformed-phase-quenched theory, i.e.,
    \begin{align}
        \nabla_p\mathcal{F}(\{p\})=\nabla_p\log\mathcal{Z}_\mathrm{pq}(\{p\})=-\big\langle \nabla_p\mathrm{Re} S_\mathrm{eff}(\{p\})\big\rangle_\mathrm{pq}^\mathrm{def}\;.
    \end{align}
    where $\langle\cdot\rangle_\mathrm{pq}^\mathrm{def}$ means that the expectation value is taken with respect to weights $|e^{-S_\mathrm{eff}(\{p\})}|$ with $S_\mathrm{eff}(\{p\})$ defined analogously as in Eq. \eqref{eq:1D} but with nonzero $\{p\}$.
    
    The success of sign optimization heavily depends on the deformation ansatz used. In this paper we use a piecewise linear approximation of a continuous contour in the form of
    \begin{align}
    \label{eq:ansatz}
        z=\varphi+if(\varphi,\dots)
    \end{align}
    with real and single-valued function $f$ that linearly interpolates between node points $x_i$ from the real line, with $i=1,2,\dots,M$, i.e.,
    \begin{align}
    \label{eq:interpolation}
    \begin{split}
        &f(\varphi,x_{j(\varphi)},x_{j(\varphi)+1},y_{j(\varphi)},y_{j(\varphi)+1})\\
        &=\frac{y_{j(\varphi)}(x_{j(\varphi)+1}-\varphi)}{x_{j(\varphi)+1}-x_{j(\varphi)}}+\frac{y_{j(\varphi)+1}(\varphi-x_{j(\varphi)})}{x_{j(\varphi)+1}-x_{j(\varphi)}}\;.
    \end{split}
    \end{align}
    The parameters to optimize are the values $y_i$ of $f(x_i)$ at the node points. The node points are kept fixed, and their spacing is chosen to be regular $x_{i+1}-x_i\equiv\Delta$, and
    \begin{align}
    \label{eq:floor}
        j(\varphi)=\mathrm{floor}\bigg[\frac{\varphi-x_1}{\Delta}\bigg]\;.
    \end{align}
    The Jacobian of such a deformation is simply
    \begin{align}
        \mathcal{J}=1+i\frac{y_{j(\varphi)+1}-y_{j(\varphi)}}{\Delta}\;.
    \end{align}
    The number of the node points $M$ defines the number of parameters to optimize. By increasing $M$ and decreasing $\Delta$ one can approximate any continuous function with arbitrary precision with such a piecewise linear function, and thus for the needed contours piecewise linearity is not really an assumption here, but continuity and being single-valued are. This ansatz was already applied in our previous work to deform the trace of a random matrix variable in Ref. \cite{Giordano:2023ppk}. One can also generalize the ansatz to include non-single-valued contours, by having an independent parametrization of the real and imaginary parts. Based on our findings, it is not needed to do so for the models considered in this paper. An alternative way to look at non-single-valued contours could be through a change of variables.
    
\section{Results on one-dimensional path integrals} 

    We studied four one-dimensional toy models in this paper; a Hubbard-, a Gross-Neveu-, a Thirring-, and Chern-Simons-like model. All of them are extreme limits of physically motivated models. For each we applied all three contour deformation methods discussed in Secs. \ref{sec:Lefschetz}, \ref{sec:holomorphic} and \ref{sec:optimization}. In this Section, we present our findings for each model. The technical details and of the calculations are rather similar for all the four models, and hence they will be described in more detail during the presentation of the first one. The finer differences and modifications will be highlighted and explained individually later on.

    \begin{figure*}[t]
        \centering
        \includegraphics[height=0.33\linewidth]{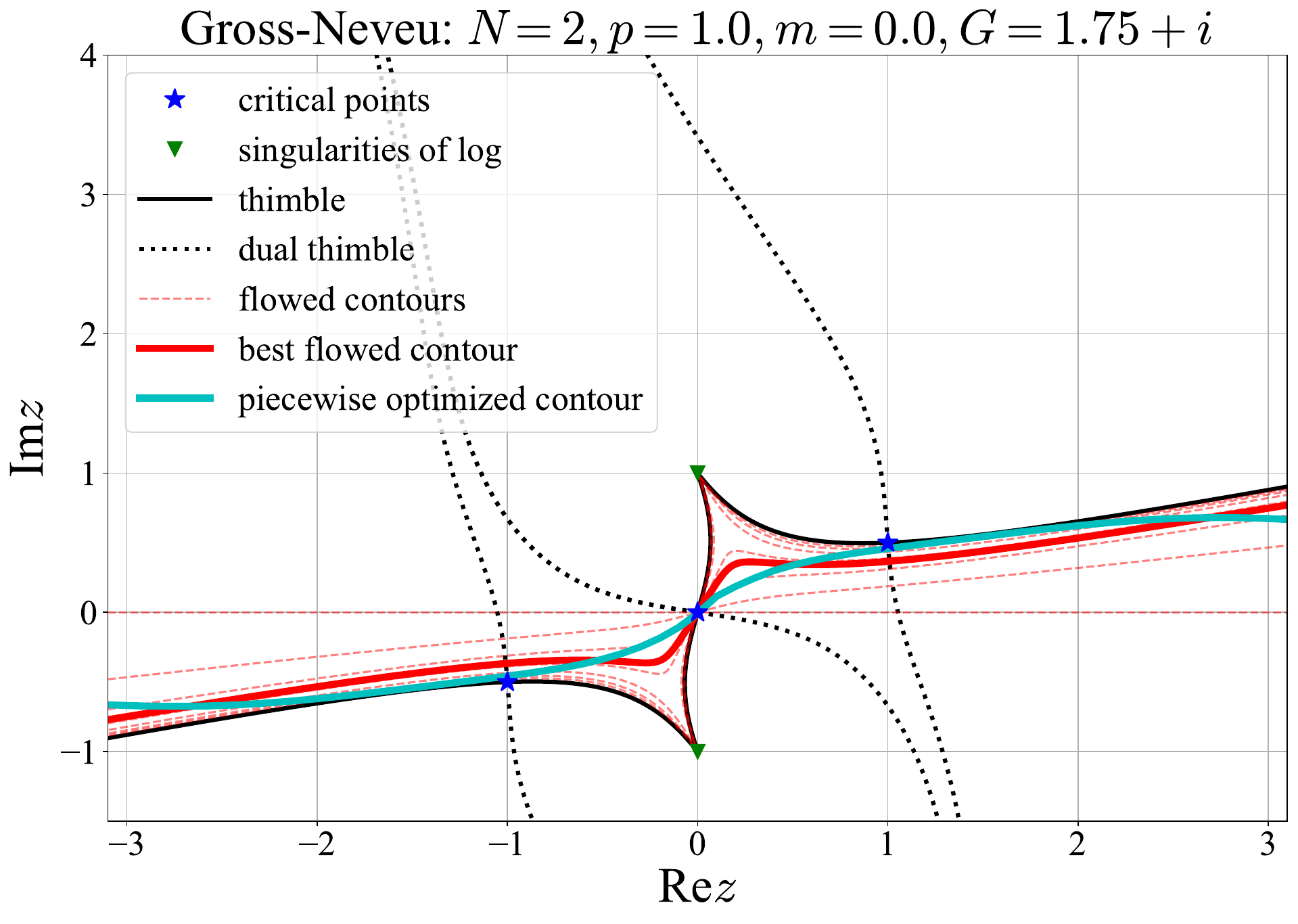}
        \includegraphics[height=0.33\linewidth]{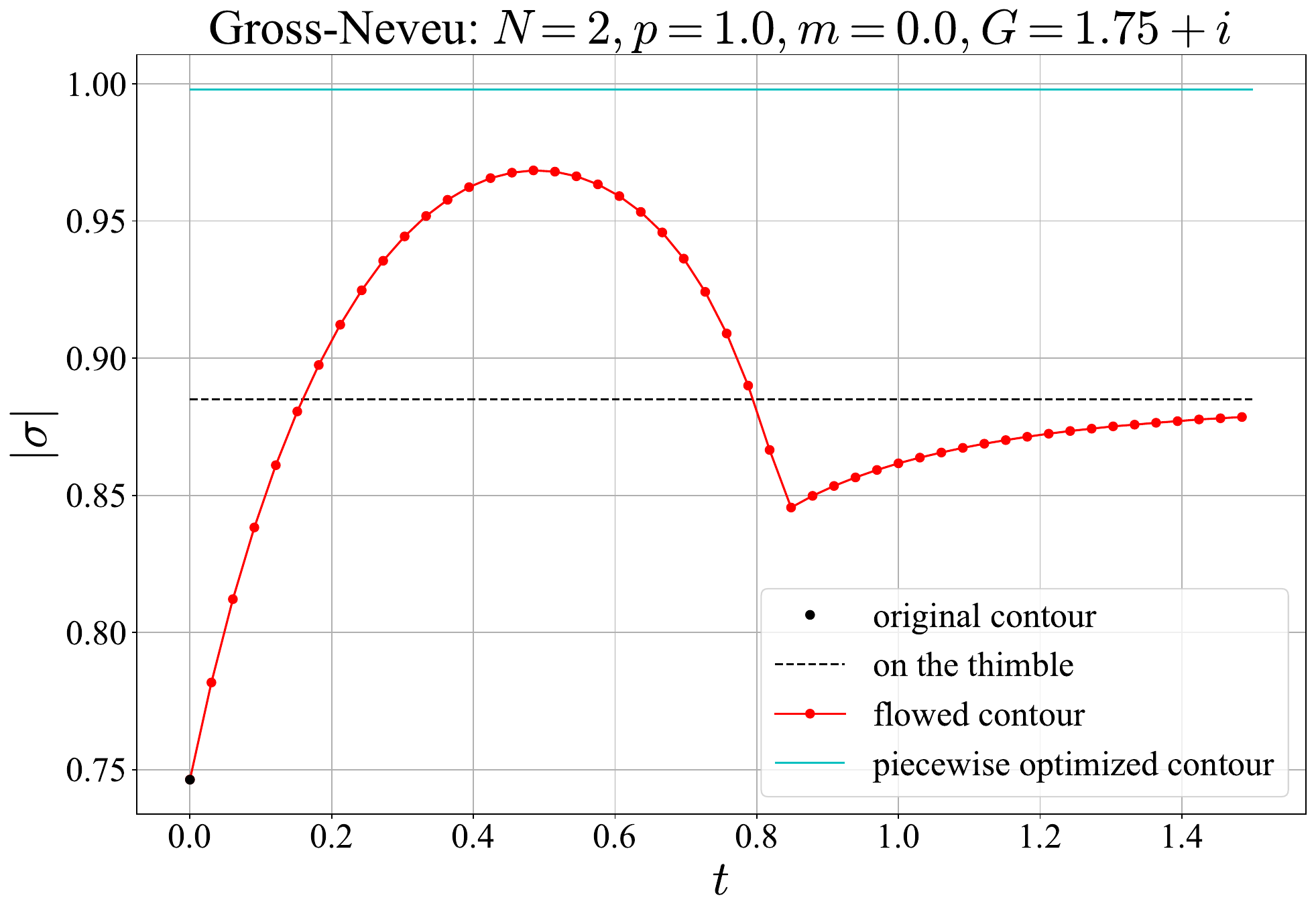}
        \caption{Left: Same as in Fig. \ref{fig:Hubbard}. but for the Gross-Neveu-like model. Right: The evolution of the average sign $\sigma$ along the holomorphic flow (red dots), and its value on the thimble (black dashed line) and on the optimized contour (solid cyan line) for the Gross-Neveu-like model at complex coupling.}
        \label{fig:GN}
    \end{figure*}    

\subsection{One-site (zero-hopping) Hubbard model}

The Hubbard model and its various extensions are prominently researched topics in condensed matter physics and it is of great interest to us as well due to its sign problem away from halffilling \cite{Wynen:2018ryx, Wynen:2020uzx, Ostmeyer:2020uov, Rodekamp:2022xpf, Gantgen:2023byf}. It describes a lattice system of strongly correlated fermions. The fermion fields at different sites are coupled through a hopping term, proportional to a hopping parameter. In the extreme limit when this hopping parameter is zero the whole lattice decouples to independent one-site Hubbard models. Upon performing a Hubbard-Stratonovich transformation, the partition function of this simple system can be expressed as a path integral over auxiliary field variables. The temporal continuum limit can be taken exactly, and all Matsubara modes can be integrated out except for the zero mode. For details of this calculation, and a semiclassical analysis of the thimble structure of the resulting integral see Ref. \cite{Tanizaki:2015rda}. The resulting exactly solvable, one-dimensional path integral with a non-compact integration domain is
    \begin{align}
    \label{eq:Hubbard_Z}
    \begin{split}
        \mathcal{Z}_\mathrm{H}&=\sqrt{\frac{1}{2\pi U}}\int_{-\infty}^\infty\mathrm{d}\varphi\,(1+e^{i\varphi+\mu+U/2})^2e^{-\varphi^2/(2U)}\;,
    \end{split}
    \end{align}
    with dimensionless chemical potential $\mu$, and interaction potential $U$ and where $\varphi$ is the zero Matsubara mode of the auxiliary bosonic field introduced by the Hubbard-Stratonovich transformation. The sign problem is severe away from half filling, i.e., $\mu/U=1/2$. 
    
    The action can be read off from Eq. \eqref{eq:Hubbard_Z} to be
    \begin{align}
    \label{eq:Hubbard_S}
        S_\mathrm{H}(z)=\frac{z^2}{2U}-2\log(1+e^{iz+\mu+U/2})\;.
    \end{align}
    There are an infinite number of critical points given by the equation
    \begin{align}
        \frac{\partial S_\mathrm{H}}{\partial z}\bigg|_{z=z_\mathrm{c}}=\frac{z_\mathrm{c}}{U}-\frac{2ie^{iz_\mathrm{c}+\mu+U/2}}{1+e^{iz_\mathrm{c}+\mu+U/2}}=0\;,
    \end{align}
    and an arbitrary number of them can be calculated numerically (e.g. with Newton-Raphson iteration). The action is also singular at 
    \begin{align}
    \label{eq:Hubbard_singular}
        z_\mathrm{s}=i(\mu+U/2)+\pi(2n+1)\;,
    \end{align}
    with $n\in\mathbb{Z}$. At these points the integrand vanishes. The thimbles and dual thimbles, the fixed flow time and optimized contours of the one-site Hubbard model are shown in Fig. \ref{fig:Hubbard} (left). The average sign $\sigma$ along the flow, on the thimbles, and on the sign optimized contour is presented in Fig. \ref{fig:Hubbard} (right).
    
   \begin{figure*}[t]
        \centering
        \includegraphics[height=0.33\linewidth]{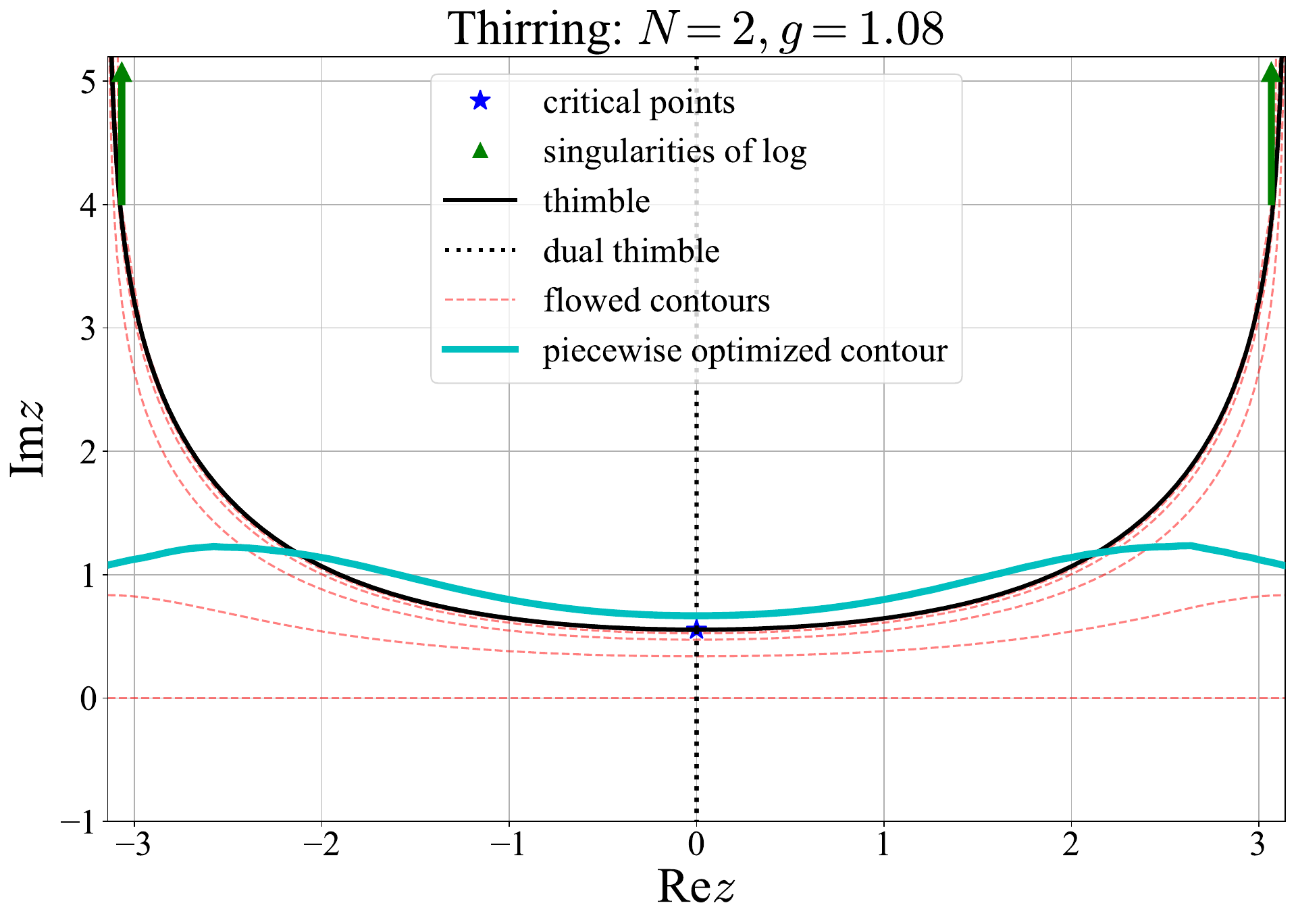}
        \includegraphics[height=0.33\linewidth]{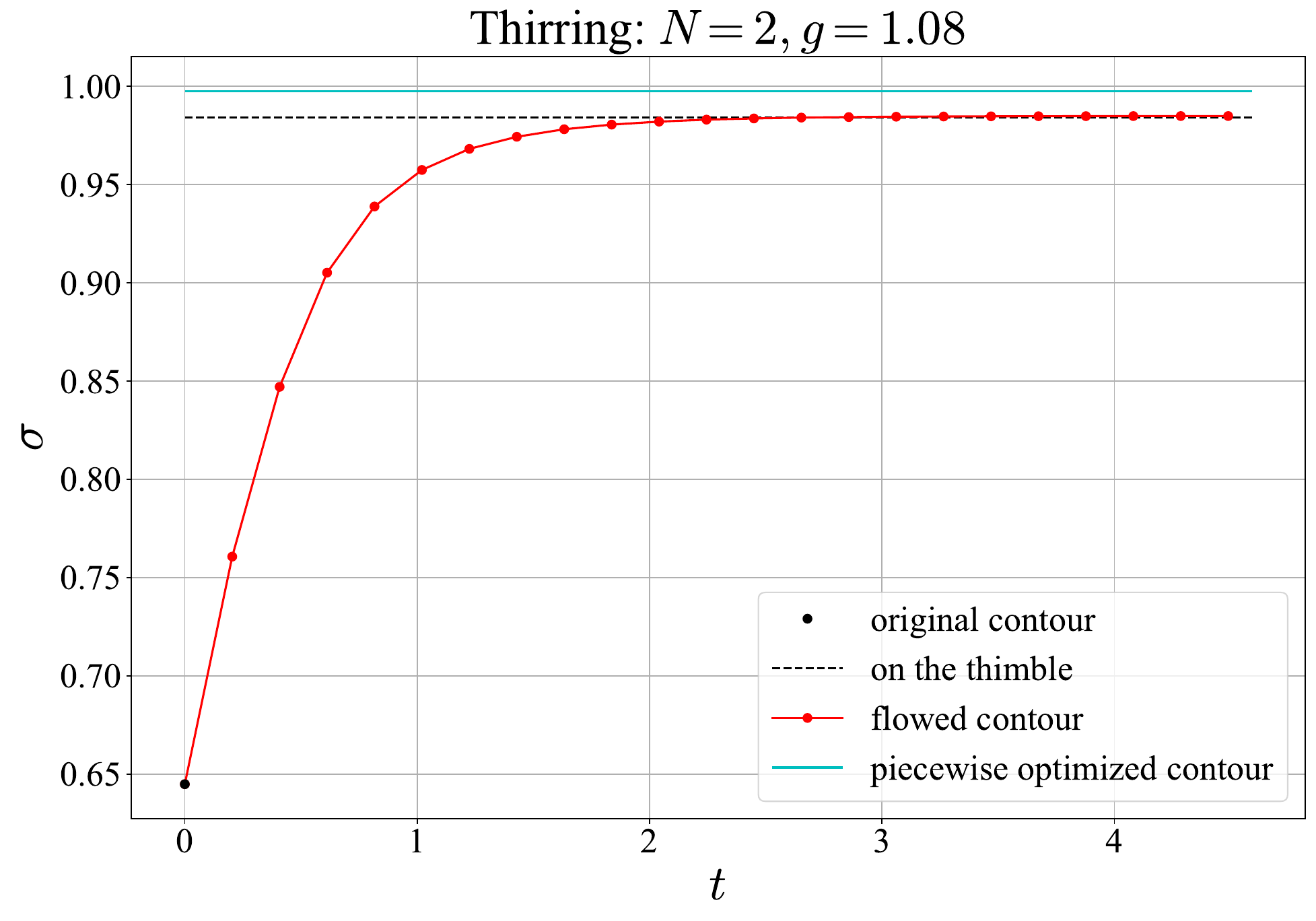}
        \caption{Left: Critical (blue stars) and singular (green wedges) points of the action of the heavy-dense Thirring model. The solid black lines are the thimbles, while the dotted black lines are the dual thimbles. The dashed red lines correspond to fixed flow time contours. In this model there is no clear separation between the thimbles and the optimally flowed contours. The solid cyan line is the optimum contour based on the sign optimization approach.
        The green arrows point toward imaginary infinity at the edges of the integration domain. Right: The evolution of the average sign $\sigma$ along the holomorphic flow (red dots), and its value on the thimble (black dashed line) and on the optimized contour (solid cyan line) for the heavy-dense Thirring model.}
        \label{fig:Thirring}
    \end{figure*}

    With the critical points $z_\mathrm{c}$ at hand, one can integrate the downward flow equation of Eq. \eqref{eq:flow_equation} backward (or the upward equation forward), setting out from each critical point $z_\mathrm{c}$, to build up the thimbles (black solid lines in Fig. \ref{fig:Hubbard} (left)). There will be an infinite number of thimbles, but typically (with a reasonable choice of parameters, for us $U=10.0$ and $\mu=0.0$) only a few will have a meaningful contribution. This is because moving away from $\mathrm{Re}z=0$ the Gaussian part of the integrand exponentially suppresses the more distant thimbles. The upward flow will tend toward the singular points $z_\mathrm{s}$, and the thimbles are going to be separated by them. The dual thimbles [black dotted lines in Fig. \ref{fig:Hubbard} (left)] were also calculated, and all of them intersect with the real line. For the integration of the differential equations an adaptive Runge-Kutta-Fehlberg solver was used. The average sign $\sigma$ was calculated from the ratio of full-to-phase-quenched partition functions, as both can be calculated precisely in this setup.

    The same solver was used to solve the coupled system of Eqs. \eqref{eq:holomorphic_flow_equation} and \eqref{eq:Jacobian_flow}, and generate contours at finite flow times. For a dense set of points picked from the real line as initial conditions the holomorphic flow equations were evolved up to such large flow times that the flowed points would approach the singular points $z_\mathrm{s}$ defined in Eq. \eqref{eq:Hubbard_singular}. And subsequently, the fixed flow time contours (red dashed lines in Fig. \ref{fig:Hubbard} (left)) were determined by interpolating all the flowed solutions to the chosen flow time values. In Fig. \ref{fig:Hubbard} (right) it can be noted that along the flow the average sign improves several orders of magnitude, and then asymptotes to its value evaluated on the thimble. However, it can be seen in the inset of Fig. \ref{fig:Hubbard} (right) that the average sign $\sigma$ does not approach its limiting value  monotonically but there exists an optimal flow time where it peaks. We attribute this behavior to the aforementioned interplay of the sign problems induced by $e^{-S}$ and the Jacobian. 

    For the sign optimization approach the Adam optimizer was used \cite{kingma2014adam}, supplemented by the gradient formulas defined by our ansatz in Eqs. \eqref{eq:ansatz}, \eqref{eq:interpolation} and \eqref{eq:floor}. The expectation values of the sign and the gradients were sampled via the Metropolis algorithm. Each optimization step included a complete Monte Carlo simulation of a slightly differently deformed phase-quenched theory. The number of parameters varied from $M=50$ to $250$. The average sign $\sigma$ was monitored throughout the optimization. Once a plateau was reached the optimization was considered finished. The contour that produced the maximal average sign $\sigma$ is the cyan solid line in the left panel of Fig. \ref{fig:Hubbard}. The optimized contour ultimately outperforms both the thimble and the optimally flowed results. By looking at the optimized contour itself in the left panel of Fig. \ref{fig:Hubbard}, one can say that it is qualitatively close to the best flowed contour, but it avoids the cusps and is much smoother in the vicinity of the log-singular points where the Jacobian of the flow is dominant.
 
\subsection{(0+1)-dimensional Gross-Neveu-like model}

    This zero-dimensional Gross-Neveu-like model was first studied in Ref. \cite{Kanazawa:2014qma} (for the original model see Ref. \cite{Gross:1974jv}). It is considered a toy model of discrete chiral symmetry breaking with four-point fermionic interactions. After carrying out a Hubbard-Stratonovich transformation the fermionic path integral can be evaluated exactly, with the remaining auxiliary field path integral
    \begin{align}
        \mathcal{Z}_\mathrm{GN}=\sqrt{\frac{N}{\pi G}}\int_{-\infty}^\infty\mathrm{d}\varphi\,\big(p^2+(m+\varphi)^2\big)^Ne^{-N\varphi^2/G}\;,
    \end{align}
    with number of flavors $N$, coupling constant $G$, momentum $p$, and fermion mass $m$. The integral domain is again noncompact similarly to the Hubbard-like integral. The sign problem is induced by making the coupling $G$ complex. This way the partition function $\mathcal{Z}_\mathrm{GN}$ becomes complex, hence the average phase as well. In order to alleviate the sign problem one wants to maximize both the real and imaginary parts of $\sigma$, thus we will plot $|\sigma|$ in our results. The thimble and dual thimble structure, and the flowed and optimized contours are plotted in the left panel of Fig. \ref{fig:GN}, and the results for $|\sigma|$ are shown in the right panel.

    The Gross-Neveu action reads
    \begin{align}
    \label{eq:S_GN}
        S_\mathrm{GN}(z)=N\frac{z^2}{G}-N\log\big(p^2+(z+m)^2\big)\;.
    \end{align}
    In contrary to the Hubbard action in Eq. $\eqref{eq:Hubbard_S}$, this one has a finite number of critical and singular points. We consider the model parameters $N=2,p=1.0,m=0.0$ and $G=1.75+i$, resulting in three critical points and two singular points of the action of Eq. \eqref{eq:S_GN}. The thimble structure of the model, the flowed and optimized contours were obtained in a similar fashion as described in the previous subsection. To each critical point a thimble is attached and with the used model parameters all of them contribute, since all of their duals intersect with the real line. The results obtained also show us a behavior already seen in the Hubbard-like integral. In the right panel of Fig. \ref{fig:GN} the magnitude of the average sign $|\sigma|$ tends toward its value on the thimble nonmonotonically developing a peak at some optimal flow time. The optimized contour outperforms both the flowed and thimble contours, and while it shows qualitative similarities to the best flowed contour it has more smoothed out shape. Note that the thimbles themselves are not single valued in the sense that their imaginary part cannot be written as a function of the real part. This property cannot be reproduced with our piecewise linear ansatz, yet, it can still produce a contour along which the sign problem is milder than on the thimbles. Furthermore, it is worth noting that the best flowed contour is single valued as well.

\subsection{Heavy-dense limit of the Thirring model}

    The heavy-dense limit of the (2+1)-dimensional Thirring model was extensively studied in Ref. \cite{Lawrence:2018mve}, where it was already shown that an appropriately chosen and optimized contour surpasses the thimbles in ameliorating the sign problem. Now we consider the same one-dimensional path integral with the holomorphic flow and our piecewise linear ansatz. The partition function is
    \begin{align}
        \mathcal{Z}_\mathrm{T}=\int_{-\pi}^\pi\mathrm{d}\varphi\;\exp\bigg(\frac{N}{g^2}\cos \varphi+i\varphi+\mu\bigg)\;,
    \end{align}
    with number of flavors $N$, coupling $g$ and chemical potential $\mu$. We use the parameter values given in Ref. \cite{Lawrence:2018mve}, i.e., $N=2$ and $g=1.08$. In contrast to the Hubbard- and Gross-Neveu-like path integrals the domain of integration is compact now.  
    The thimbles, flowed and optimized contours are presented in the left panel of Fig. \ref{fig:Thirring}.
    
    The action,
    \begin{align}
        S_\mathrm{T}(z)=-\frac{N}{g^2}\cos z-iz\;,
    \end{align}
    has a single critical point at $z_\mathrm{c}=i\,\mathrm{arsinh}(g^2/N_\mathrm{f})+2\pi n$, with $n\in\mathbb{Z}$ (only $n=0$ stays in $[-\pi,\pi]$), and becomes singular on the edges of the integration domain. This will result in thimbles stretching out to imaginary infinity at $\mathrm{Re}z=\pm\pi$. The average sign along the flow is shown in the right panel of Fig. \ref{fig:Thirring}. Again, as expected from Ref.~\cite{Lawrence:2018mve}, the optimized contour outpaces the thimbles. The flow results supersedes the thimble only slightly, with their maximum difference being $\mathcal{O}(10^{-4})$.

\subsection{(0+1)-dimensional Chern-Simons-like theory with fermions}

\begin{figure*}
        \centering
        \includegraphics[height=0.33\linewidth]{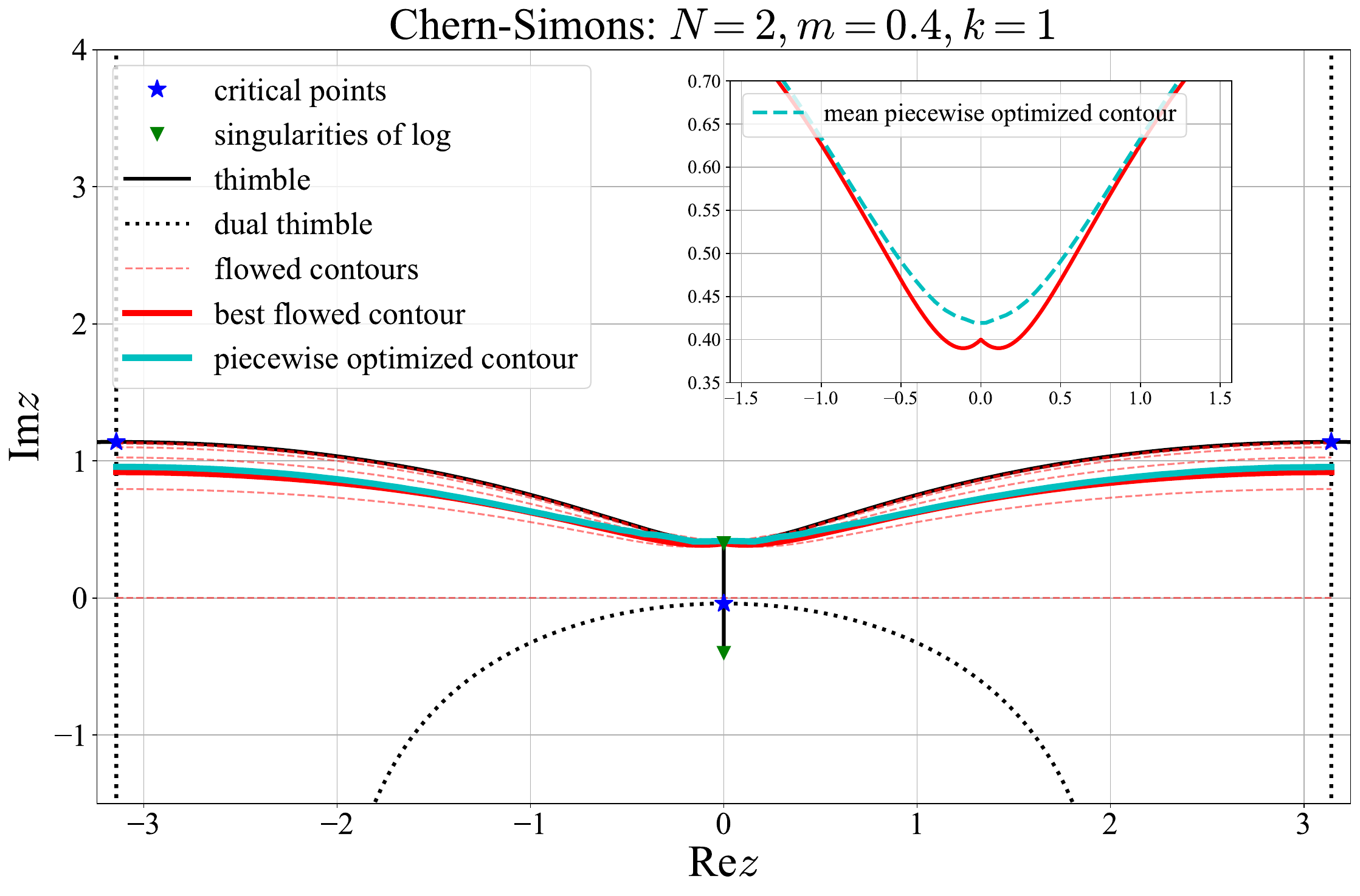}
        \includegraphics[height=0.33\linewidth]{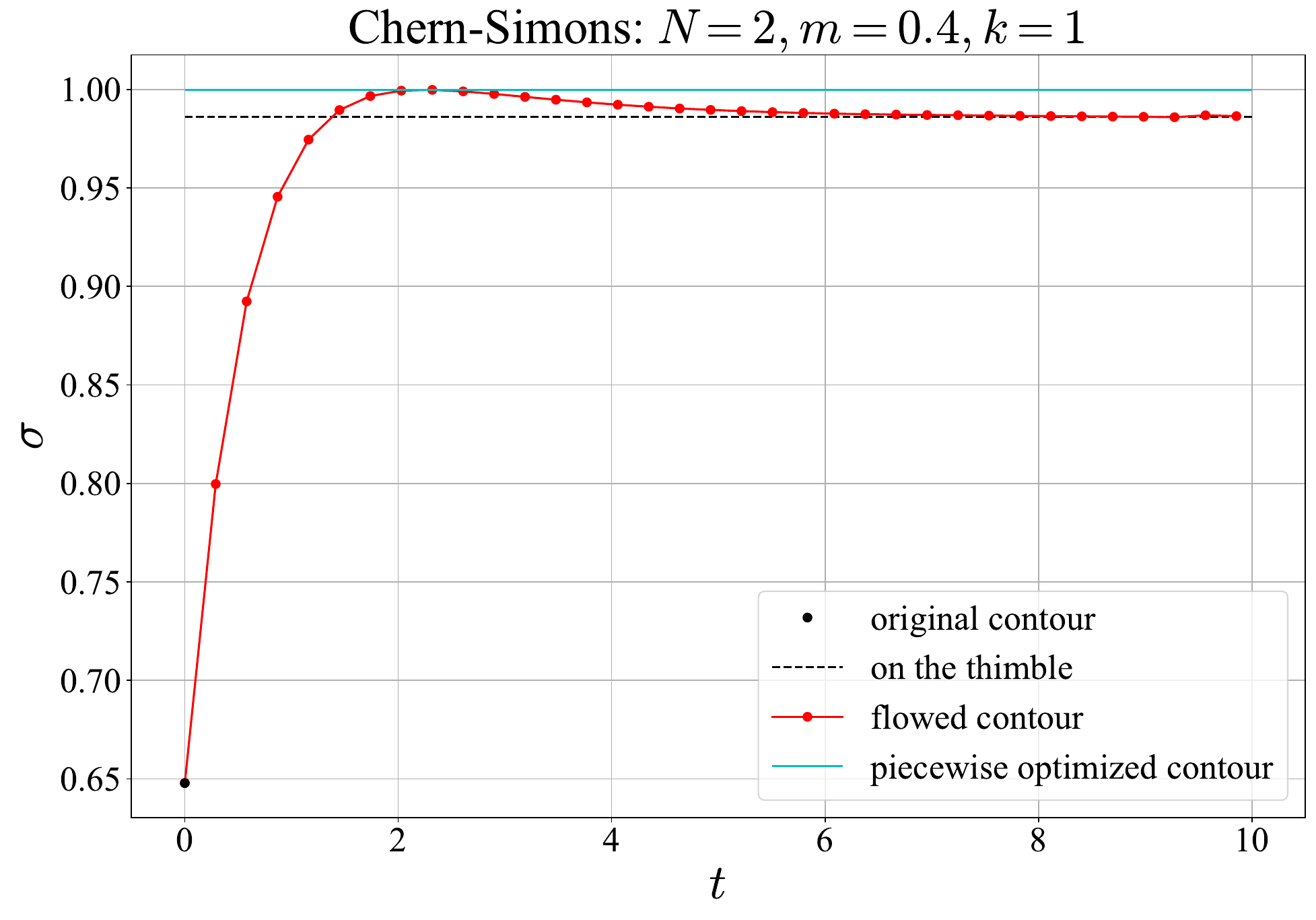}
        \caption{Left panel: Critical (blue stars) and singular (green wedges) points of the action of the Chern-Simmons-like model. The solid black lines are the thimbles, while the dotted black lines are the dual thimbles. The dashed red lines correspond to fixed flow time contours, while the solid red line is the one flowed contour that produces the maximal average sign. The solid cyan line is the optimum contour based on the sign optimization approach. Right panel: The evolution of the average sign $\sigma$ along the holomorphic flow (red dots), and its value on the thimble (black dashed line) and on the optimized contour (solid cyan line) for the Chern-Simmons-like model.}
        \label{fig:CS}
    \end{figure*}

    Last, we consider a Chern-Simons-like theory with fermions included \cite{Dunne:1996yb}, with its thimbles structure studied also in Ref. \cite{Kanazawa:2014qma}. In this model the sign problem originates from a topological term analogous to the $\theta$ term in QCD. Upon gauge fixing, and integrating out the Grassmann fields the partition function becomes
    \begin{align}
        \mathcal{Z}_\mathrm{CS}=\frac{e^{-ik\pi}}{2^N}\int_{-\pi}^\pi\mathrm{d}\varphi\;(\cosh m-\cos\varphi)^Ne^{ik\varphi}\;,
    \end{align}
    where $k\in\mathbb{Z}$ is the topological charge, $N$ is the number of fermion flavors, and $m$ is the fermion mass. The integration domain, similarly to the Thirring-like model, is compact. A sign problem will be present in the model for any $k\neq0$. Our results on the contours and the evolution of the average sign along the flow is shown in Fig. \ref{fig:CS}.

    The action of the model is
    \begin{align}
        S_\mathrm{CS}(z)=-N\log(\cosh m-\cos z)-ikz\;.
    \end{align}
    Our interest is the scenario when $k<N$, as for $k>N$ the partition function is zero, while for $k=N$ is a special case where the critical points are at imaginary infinity. With the used model parameters, $N=2,m=0.4$ and $k=1$, three critical points of the action are present in the range $[-\pi,\pi]$, and a complex conjugate pair of singular points. However, one of the dual thimbles (the black dotted lines in the left panel of Fig. \ref{fig:CS}) does not pierce the real line, and hence its associated thimble does not contribute. The inset of the left panel of Fig. \ref{fig:CS} shows that the best flowed contour develops a cusp in the vicinity of the singular point of the action. The sign optimized contour remains smooth. Actually, the middle part of the contour does not have a significant contribution, as it fluctuates prominently in the course of the optimization procedure even when the average sign $\sigma$ has already reached its optimal value. The dashed cyan line in the inset is the mean of the optimal contours, i.e., the average of $\mathcal{O}(1000)$ contours already optimized to have maximal average phase. In the right panel of Fig. \ref{fig:CS} the nonmonotonic nature of the flowed average sign is visible in the case of the Chern-Simons-like model as well. The best flowed and the sign optimized contours both produce $\sigma\approx1$ (at these model parameters), and are barely distinguishable from each other, i.e., their corresponding $\sigma$ values differ only in their fourth decimals. 
    
\section{Discussion:}
    We studied four different oscillatory integrals, all of which were derived as an extreme (one-site) limits of some physically interesting model: a Hubbard-like-, a Gross-Neveu-like-, a Thirring-like- and a Chern-Simmons-like integral.  
    We calculated the Lefschetz thimbles, as well as the integration contours generated with the holomorphic flow equations, and the numerically optimized contours with the largest average phase $\sigma$ for
    all of these models. In every case, the numerically optimized contours could provide a significantly larger $\sigma$ value than what was achievable on the thimbles. Hence, what was noted for one example in Ref.~\cite{Lawrence:2018mve} seems to be a generic feature. We observed that the convergence of the holomorphic flow method to the value of $\sigma$ achievable on the thimbles is not monotonic, implying the existence of an optimal flow time, where the finite flow time manifold outperforms the thimbles. An encouraging aspect of our findings is that
    the numerically optimized integration contours do not typically show the kind of behavior that makes dealing with the thimbles numerically difficult: we observed no cusps near the zeros of the fermion determinant and no divergence to infinity for these contours. 

    A clear limitation of our work is that we only considered one-dimensional integrals. For the case of zero-hopping Hubbard model, where the partition function is simply the one-site partition function to the power of the number of sites, the values of $\sigma$ we found on the different integration contours would give 
    for the severity of the sign problem: 
    \begin{equation}
    \sigma (N\ \mathrm{sites}, 0\ \mathrm{hopping}) = \sigma(1\ \mathrm{site}, 0\ \mathrm{hopping})^{N}.
    \end{equation}
    Thus, as $N$ is increased, even a small reduction in the $\sigma$ value for the one-site integral becomes very significant. How the difference between the numerically optimal contours and the thimbles changes for the case of nonzero hopping and many degrees of freedom cannot be inferred from this work, but is an interesting topic for future research.
    
    While it is not practical for very high dimensional integrals, the piecewise linear ansatz we used for our toy integrals can 
    be straightforwardly generalized to $N$-dimensional integrals as well. In this case one has to specify $N$ functions denoted $F_i$ with $i=1,2,\dots,N$ (the imaginary parts of the integration variables) of $N$ real variables (the real parts of the integration variables) denoted $x_j$ with $j=1,2,\dots,N$. Introducing an $N$-dimensional hypercubic grid of size $M$ in each direction, with points denoted $(x_j)_k \equiv x_{jk}$ with $j=1,2,\dots,N$ and $k=1,2,\dots,M$ and specifying the function values $F_i(x_{jk})$ requires $N \times M^N$ parameters. The analog of the 
    piecewise linear interpolation is achieved by splitting the $N$-cube cells into $N!$ simplices. Each simplex has $N+1$ vertices, where the 
    function values $F_i(x_{jk})$ are known. This data can be used to define a hyperplane, functioning as a local linear interpolant. This way
    the piecewise linear ansatz used in this work can be generalized into arbitrary dimensions. However, for large $N$ this approach is not practically usable for the same reason that functional evaluations of the integrand on an $N$-dimensional hypercubic grid is an inefficient way
    to evaluate nonoscillating $N$-dimensional integrals: the number of grid points scales exponentially with the number of dimensions $N$. Thus, to have an efficient implementation in higher dimensions, a notion of importance sampling has to be introduced already in the locally linear ansatz. E.g. generating $n$ points with the phase-quenched probability density and using Delaunay triangulation to construct a piecewise linear ansatz would be a slightly better method. However, in this case the triangulation steps scales like $n^{N/2}$ (for even $N$) which is still computationally heavy. Thus, while this approach can be used for small $N$,    
    for higher dimensional
    integrals, other approaches should be used to find numerically optimal contours. 
    
    In multiple dimensions one should also be careful that the Jacobian determinant is nonzero. This issue was discussed by us in the context of a chiral random matrix model in Ref. \cite{Giordano:2023ppk}. Furthermore, higher dimensional integrals have important features (such as spacetime and internal symmetries) that cannot be mimicked with the one-dimensional toy examples of this paper. Numerical optimization methods in these models should certainly take these features into account. This is beyond the scope of this work. 

\section*{Acknowledgments} This work was supported by the Hungarian National
Research, Development and Innovation Office under
Project No. FK 147164. D. P. is supported by a
Fulbright Program grant sponsored by the Bureau of
Educational and Cultural Affairs of the United States
Department of State and administered by the Institute of
International Education and the Hungarian-American
Commission for Educational Exchange, with additional
support by the EKÖP-24 University Excellence
Scholarship Program of the Ministry for Culture and
Innovation from the source of the National Research,
Development and Innovation Fund. A. P. acknowledges
support of the Momentum Programme of the Hungarian
Academy of Sciences (grant LP2025-13/2025). This work
was also supported by NKFIH Grants No. TKP2021-
NKTA-64, NKKP Excellence No. 151482.

\section*{Data Availability} The data that support the findings of this article are not
publicly available. The data are available from the authors
upon reasonable request.
\providecommand{\noopsort}[1]{}\providecommand{\singleletter}[1]{#1}%
%


\end{document}